# Transport properties of  $\beta$-Ga$_2$O$_3$ Nanoparticles embedded in Nb thin films

L.S.Vaidhyanathan[a*] , M.P. Srinivasan[b] , P. Chandra Mohan[b] , D.K. Baisnab[a]  R. Mythili[c] and   M.P. Janawadkar[a]

[a]Condensed Matter Physics Division, Indira Gandhi Centre for Atomic Research, Kalpakkam, India.

[b] Water and Steam Chemistry Division, Bhabha Atomic Research Centre Facilities, Kalpakkam, India

c Microscopy & Thermo-Physical Property Division,, Indira Gandhi Centre for Atomic Research, Kalpakkam, India.

## Abstract

The origin of ferromagnetism in nanoparticles of nonmagnetic oxides is an interesting area of research. In the present work, transport properties of niobium thin films, with $\beta$-Ga$_2$O$_3$ nanoparticles embedded within them, are presented. Nanoparticles of $\beta$-Ga$_2$O$_3$ embedded in a Nb matrix were prepared at room temperature by  radio frequency  co-sputtering technique on Si (100) and glass substrates held at room temperature.   The thin films deposited on Si substrates were subjected to  Ar annealing at a temperature range of 600-650 C for 1 hour. Films were characterized by X-ray diffraction (XRD), Micro-Raman and elemental identification was performed  with an Energy Dispersive X-ray Spectroscopy (EDS). Transport measurements were performed down to liquid helium temperatures by four-probe contact technique, showed characteristics analogous to those observed in the context of a Kondo system. A comparison of the experimental data with the theoretical formalism of Kondo and Hamann is presented. It is suggested that this behavior arises from the existence of magnetic moments associated with the oxygen vacancy defects in the nanoparticles of the nonmagnetic oxide Ga$_2$O$_3$.

* Corresponding author: Telefax: 0091-44-27480081;
Email address: lsv@igcar.gov.in (L.S. Vaidhyanathan)

## 1. Introduction

The Kondo effect which is characterized by logT behaviour in the temperature dependence of electrical resistivity at low temperatures, has been the subject of extensive theoretical and experimental research. It has been studied in both crystalline [1] and amorphous [2] systems with the introduction of magnetic impurities carrying a magnetic moment. Recent experiments have lent credence to the existence of a new fundamental state of magnetism, viz "Quantum Spin Liquid" [3] and Anderson, who first proposed this concept, has explored its relevance in the context of High temperature Superconductivity [4]. The topic related to the interaction of superconductivity with magnetism is of great theoretical and experimental interest and there are many interesting problems which are at the frontiers of current research. The experimental investigations range from studies on proximity effect in superconductor (S)-ferromagnet (F) bilayers, SFS trilayers, spin injection into the superconductors, $\pi$ type superconductivity etc. On the theoretical side, in SF structures existence of spatial modulation of the superconducting pair potential is envisaged and exotic phenomenon of spin triplet superconductivity in SFS system is anticipated. Room temperature ferromagnetism has been recently observed in nanoparticles of several nonmagnetic oxides [5] and the origin of ferromagnetism is still a matter of debate. $Ga_2O_3$ is an exclusive deep ultraviolet transparent conducting oxide and it has interesting electrical conductivity characteristics, such as the high electrical conductivity of ~ 40 S cm$^{-1}$ induced by tin doping [6]. An insulator to metal transition with a conductivity enhancement of seven orders of magnitude [7] has been observed in amorphous non-stoichiometric Gallium oxide thin films and has been explained as arising due to heterogeneous solid state reaction.

In view of this, it would be interesting to investigate the effect of $Ga_2O_3$ on the superconducting properties of materials such as Nb. This paper reports the investigations on the electrical transport properties of $Ga_2O_3$/Nb thin films, especially after post annealing the film under reducing conditions at elevated temperatures, and the experimental data has been analyzed using the theoretical formalism proposed by Kondo and Hamann [8].

## 2. Experiments

Nanoparticles of β-Ga$_2$O$_3$ embedded in a Nb matrix were prepared at room temperature by radio frequency co-sputtering technique after establishing a base pressure of 1.3 x 10$^{-7}$ mbar. Nb target of 99.99% purity and ultra high pure Ar gas (99.9995%) were used to deposit the thin films. Ga$_2$O$_3$ nanoparticles were formed in a Nb matrix by co-sputtering a Nb target (10 cm diameter) along with 2 cm diameter target of Ga$_2$O$_3$ to deposit the thin film on Si (100) and glass substrates held at room temperature. The composite films of thickness of 200 nm were deposited in the presence of ultra high pure Ar gas at a pressure of 8 mTorr with incident RF power of 90 W. Electrical contacts to the thin films were provided in the standard four probe contact geometry using UV photolithography. The thin films deposited on Si substrates were subjected to annealing in argon atmosphere by keeping the films in ultra high pure Argon gas at a pressure of 15 mbar at two different temperatures (a) T = 600 C for 30 minutes (batch – I) and (b) T = 650 C for 60 minutes (batch – II). The crystal structure and orientation of the thin films were characterized by X-ray diffraction (XRD) measurements in a STOE diffractometer using CuK$_\alpha$ radiation. Micro-Raman study is a sensitive technique to probe the local atomic environments and here Confocal Dispersive Raman microscopy with Argon ion laser (514.5nm) was used for acquiring the spectra from the sample surface in the back scattering mode. The elemental identification of Nb-Ga$_2$O$_3$ thin films on glass substrates was carried out using Philips XL- 30 SEM fitted with an Energy Dispersive X-ray Spectroscopy (EDS). Transport measurements were performed down to liquid helium temperatures by four-probe contact technique using indium soldering to establish the contacts.

## 3. Results and Discussion

Fig. 1a shows the room temperature GIXRD pattern of Nb-Ga$_2$O$_3$ film using CuK$_\alpha$ radiation showing (110), (200) and (211) Bragg peaks corresponding to only Nb. No lines corresponding to the Ga$_2$O$_3$ phase could be detected since the volume fraction of Ga$_2$O$_3$ in the film estimated from the deposition rates is less than the detection limit for XRD of 5 % Fig. 1b shows the room temperature Raman spectra recorded from Nb-Ga$_2$O$_3$ thin film prepared on borosilicate glass substrate. The peak at 200 cm$^{-1}$ arises from the different vibrational modes of β- Ga$_2$O$_3$ and the peak at 806 cm$^{-1}$ is ascribed to symmetric stretching motions of silica tetrahedra with oxygen in the borosilicate glass [9]. It is generally seen in literature that the detection of Raman spectrum from a metal (like Nb) is beset with lot of difficulties. Fig. 1c shows the EDS spectra showing the elemental identity of Nb and Ga through their characteristic X-ray energies. The other element such as Ca is also observed due to glass substrate.

The temperature variation of electrical resistivity ρ(T) of Ar annealed Nb-Ga$_2$O$_3$ thin film (batch –I) is shown in Fig. 2 for the temperature range between 5 K to 210 K. The electrical resistivity has a value of ~ 80 μΩ cm at 210 K, increases with decrease in temperature, exhibits a broad maximum around 50 K and then decreases with further decrease in temperature. The top inset shows the schematics of film deposition and measurement performed. The inset at the bottom shows the variation of ρ(T) of the as deposited Nb-Ga$_2$O$_3$ thin film which exhibits a normal metallic behaviour with decrease in temperature but is not superconducting down to 4.2 K. An rough estimate of the particle size of Nb using Scherrer formula turns out to be 10 nm. The nanostructured Nb thin films of size of ~ 10 nm does not show superconducting behaviour. Added to it, the Presence of Ga$_2$O$_3$ nanoparticles dispersed in niobium may have the potential to radically alter the conduction behaviour as well as superconductivity.

Similarly, the temperature dependence of the electrical resistivity ρ (T) of batch – II sample of Ar annealed Nb-Ga$_2$O$_3$ thin film is shown in Fig. 3. It may be noted that ρ(T) for this specimen also increases with decrease in temperature with an eventual saturation of ρ(T) at temperatures between 35K and 4.2K. Our analysis of the experimental data is based on the observation of magnetism in nanoparticles of several non-magnetic oxides such as Al$_2$O$_3$, SnO$_2$, In$_2$O$_3$ etc [5]

A characteristic feature of Kondo system is the observation of a resistance minimum with a logarithmic divergence of resistivity at lower temperatures. It is well known that when the exchange interaction between the magnetic moment of the impurity and the conduction electron spins is antiferromagnetic, the contribution to the electrical resistivity arising from spin flip scattering varies as lnT.

$$\rho(T) = (\rho_O + \rho_O^\infty) - a_K \ln T \qquad (1)$$

here, $\rho_o$ is the contribution from the scattering of the conduction electrons with lattice defects, while $\rho_o^\infty$ is that due to non spin flip scattering. The second term accounts for Kondo effect viz., spin flip scattering of the conduction electrons by the magnetic moments

The temperature variation of electrical resistivity $\rho(T)$ of Ar annealed Nb-Ga$_2$O$_3$ thin film (batch –I) is fitted with eqn.1. A good fit to the data (Fig. 2) is obtained for temperatures above 75 K and is shown as a solid line in Fig. 2 with the fitting parameters ($\rho_o + \rho_o^\infty$) = 90 µΩ cm and $a_K$ = 2.2 µΩ cm. (Table I)

The temperature variation of electrical resistivity $\rho(T)$ for Nb-Ga$_2$O$_3$ thin film (batch –II), which was subjected to Argon annealing for a longer duration of time, was also fitted to eqn.1 and the results are shown in Fig. 3 (inset). A good fit to the data could, however, be obtained only in the temperature range above 235 K to room temperature and is shown as a solid line with the fitting parameters ($\rho_o + \rho_o^\infty$) = 880 µΩ cm and $a_K$ = 130 µΩ cm (Table I).

It may be noted that at low temperatures, the electrical resistivity $\rho$ deviates from the expected lnT behaviour and begins to saturate at lower temperatures. To further analyze the resistivity, the experimental data of the thin films of batch –II is fitted to the expression given by Hamann [8]

$$\rho_{sd} = \frac{\rho_O}{2}\left(1 - \frac{\ln(T/T_{KO})}{\left(\ln^2(T/T_{KO}) + S(S+1)\pi^2\right)^{1/2}}\right) \qquad (2)$$

Hamann has shown that, similar to the anomalous conduction electron scattering enunciated by Kondo from the s-d exchange model, the Anderson extra orbital dilute alloy model is also of value in accounting for the experimentally observed resistivity behaviour. In eq.2. $T_{Ko}$ is the Kondo temperature and S is the impurity spin value and $\rho_0$ is the unitarity limit. Resistivity $\rho$ of the sample is represented by $\rho = \rho_{sd} + <B>$, where $<B>$ is the temperature independent contribution to the resistivity to account for the potential scattering background. The parameters $T_{Ko}$, $\rho_o$ and $<B>$ are varied to fit the experimental values for $\rho$ as per eqn.2 are shown in Fig. 3 and the values of the fitting parameters have been tabulated in Table I .

One of the characteristic features of Kondo system is the occurrence of plateau of resistivity at low temperatures. Typically saturation starts at ~ 0.1 $T_{Ko}$. If $T_{Ko}$ ~ 300 K, saturation will roughly set in at T ~ 30 K. One can see from Fig. 3 that such a plateau of resistivity is seen at T ≤ 30 K. In order to understand the low temperature saturation of resistivity as seen in Fig. 3 for the Ar annealed Nb-$Ga_2O_3$ thin film sample of batch-II, the resistivity measured for the as deposited Nb-$Ga_2O_3$ (without any annealing) was subtracted and plotted in Fig. 4. A very good fit for the entire temperature range to the experimental data was obtained and the fitting parameters are shown in Table I.

Both Kondo and Hamann theories make definite predictions for the temperature dependence of electrical resistivity at low temperatures. The Kondo effect has been experimentally observed in several systems which also include amorphous alloys. For example, the introduction of the magnetic impurities such as Cr into an amorphous alloy of $Ni_{41}Pd_{41}B_{18}$ was seen to affect the transport properties and revealed a characteristic logarithmic temperature dependence of resistivity [10]. It is also customary in these studies to separate out the magnetic contribution from the parent compound to study the temperature dependence of resistivity [11]. However, the system Nb-$Ga_2O_3$ is interesting because both Nb and $Ga_2O_3$ are known to be nonmagnetic in bulk. However, the possibility of magnetism in several nonmagnetic oxides is being actively investigated in the current literature [12-13] with a view to understand the origin of magnetism in such oxides and its influence on the physical properties. It is speculated that the origin of magnetism in such nonmagnetic oxides may arise from the localized electron spin moments ensuing from the oxygen vacancies at the surfaces of nanoparticles of these materials. In the present work, transport properties of as deposited Nb-$Ga_2O_3$ thin films exhibited the normal metallic behaviour from room temperature down to ~ 30 K while Ar annealed thin films of Nb-$Ga_2O_3$ exhibited a non-metallic behaviour and revealed a characteristic lnT temperature dependence at low temperatures probably due to the existence of localized magnetic moments associated with the oxygen vacancy defects. Indeed, the experimentally observed behaviour of the temperature dependence of electrical resistivity showing Kondo-like anomaly lends credence to the existence of magnetic moments in this system. Although challenging, it would be of interest to study the temperature dependence of magnetic susceptibility to throw light on presence of magnetic moments in this system.

## 4. Conclusion

In summary, $\beta$-$Ga_2O_3$ nanoparticles embedded in Nb thin films were deposited by RF magnetron sputtering. For Nb/$Ga_2O_3$ thin films annealed in Argon atmosphere, the temperature dependence of electrical resistivity at low temperatures showed characteristics analogous to those observed in a Kondo system. The experimental data have been analyzed using the theoretical formalism proposed by Kondo and Hamann, which predict a characteristic logarithmic temperature dependence of resistivity at low temperatures. The experimentally observed behavior of electrical resistivity is in broad agreement with these predictions, pointing to the existence of magnetic moments at oxygen vacancy defects in the nanoparticles of these nonmagnetic oxides.

**Acknowledgements**

The authors would like to thank Dr V. Sridharan for the help in providing the $Ga_2O_3$ sample. Thanks are due to Ms Kalavathi for help in X-ray measurements. We thank

Dr C.S. Sundar, Director MSG, for his interest in our work. We acknowledge the support received from IGCAR during the execution of this work.

**Table 1**: Fit parameters of temperature variation of electrical resistivity $\rho(T)$ of Ar annealed Nb- $Ga_2O_3$ thin films

| Ar annealed Nb-$Ga_2O_3$ thin film | Kondo theory | Hamann theory |
|---|---|---|
| Bartch-I | $(\rho_o + \rho_o^{\infty}) = 90\ \mu\Omega$ cm<br>$a_K = 2.2\ \mu\Omega$ cm | |
| Batch-II | $(\rho_o + \rho_o^{\infty}) = 880\ \mu\Omega$ cm<br>$a_K = 130\ \mu\Omega$ cm | $\rho_o = 90\ \mu\Omega$ cm<br>$T_{Ko} = 299$ K<br>$\langle B \rangle = 70\ \mu\Omega$ cm |
| $\Delta\rho$ ($\rho$[Ar annealed Nb-$Ga_2O_3$ : batch II] – $\rho$[Nb-$Ga_2O_3$ as deposited]) | | $\rho_o = 130\ \mu\Omega$ cm<br>$T_{Ko} = 315$ K<br>$\langle B \rangle = 9.7\ \mu\Omega$ cm |

**Figure Captions**

**Fig. 1.** (a) Room temperature GIXRD pattern of Nb-$Ga_2O_3$ film; (b) Room temperature Raman spectra recorded from Nb-$Ga_2O_3$ thin film prepared on borosilicate glass substrate; (c) EDS spectra of Nb-$Ga_2O_3$ thin film prepared on borosilicate glass substrate.

**Fig. 2.** The temperature variation of electrical resistivity $\rho(T)$ of Ar annealed Nb-$Ga_2O_3$ thin film (batch –I). The linear variation with ln(T) at temperatures above 75K (shown by the solid line) indicates Kondo –like behaviour. Inset shows the variation of $\rho(T)$ of the as deposited Nb-$Ga_2O_3$ thin film (without any annealing); top inset shows the schematics of film deposition and measurement performed.

**Fig. 3.** Temperature variation of electrical resistivity $\rho(T)$ of Ar annealed Nb-$Ga_2O_3$ thin film (batch –II). Line denotes a fit due to Hamann's theory. Inset shows a fit (solid line) to a logarithmic temperature dependence.

**Fig. 4.** Comparison of Hamann's theory (solid line) with the $\rho(T)$ data of Ar annealed Nb-$Ga_2O_3$ thin film (batch – II), after subtracting the electrical resistivity of the as deposited Nb-$Ga_2O_3$ thin film.

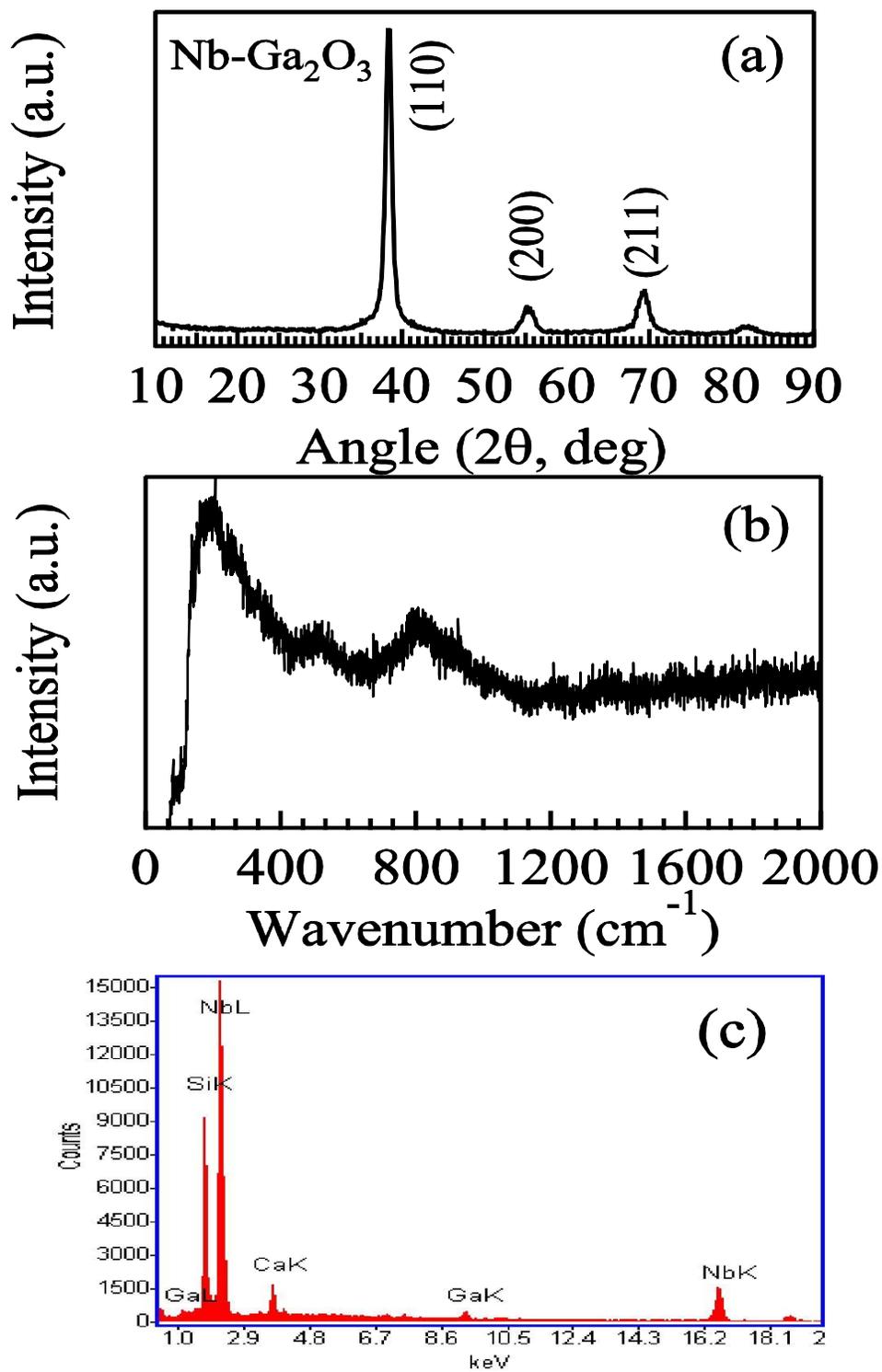

Fig.1.

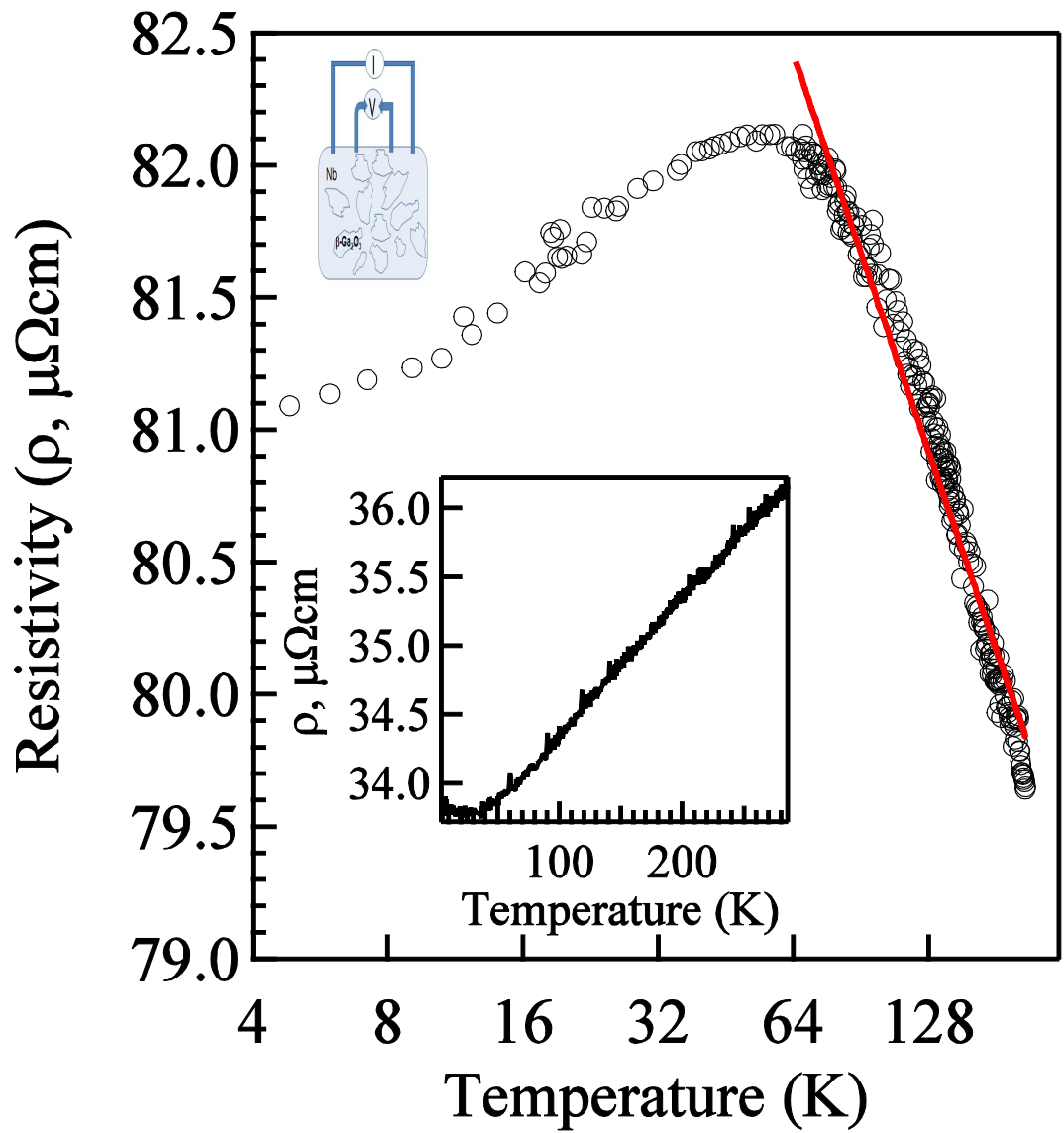

Fig.2.

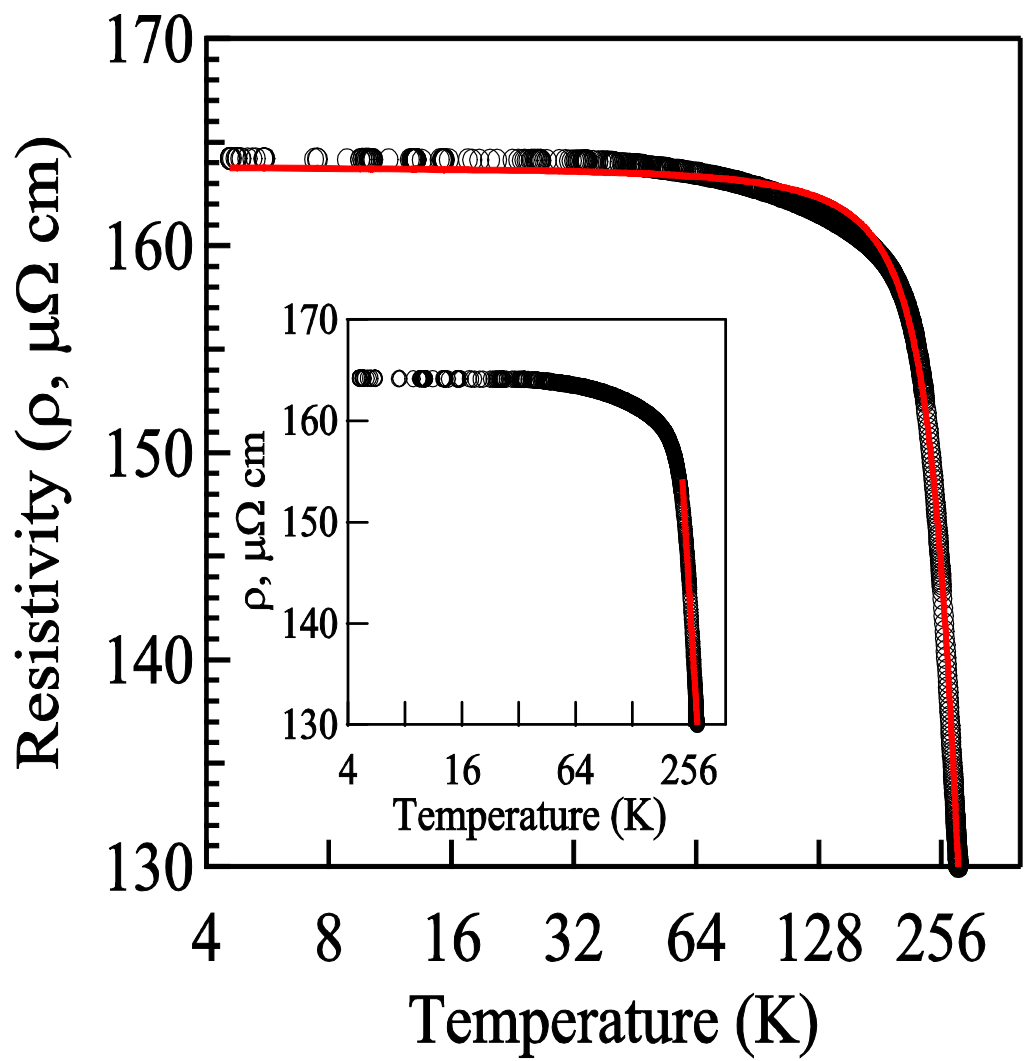

Fig.3.

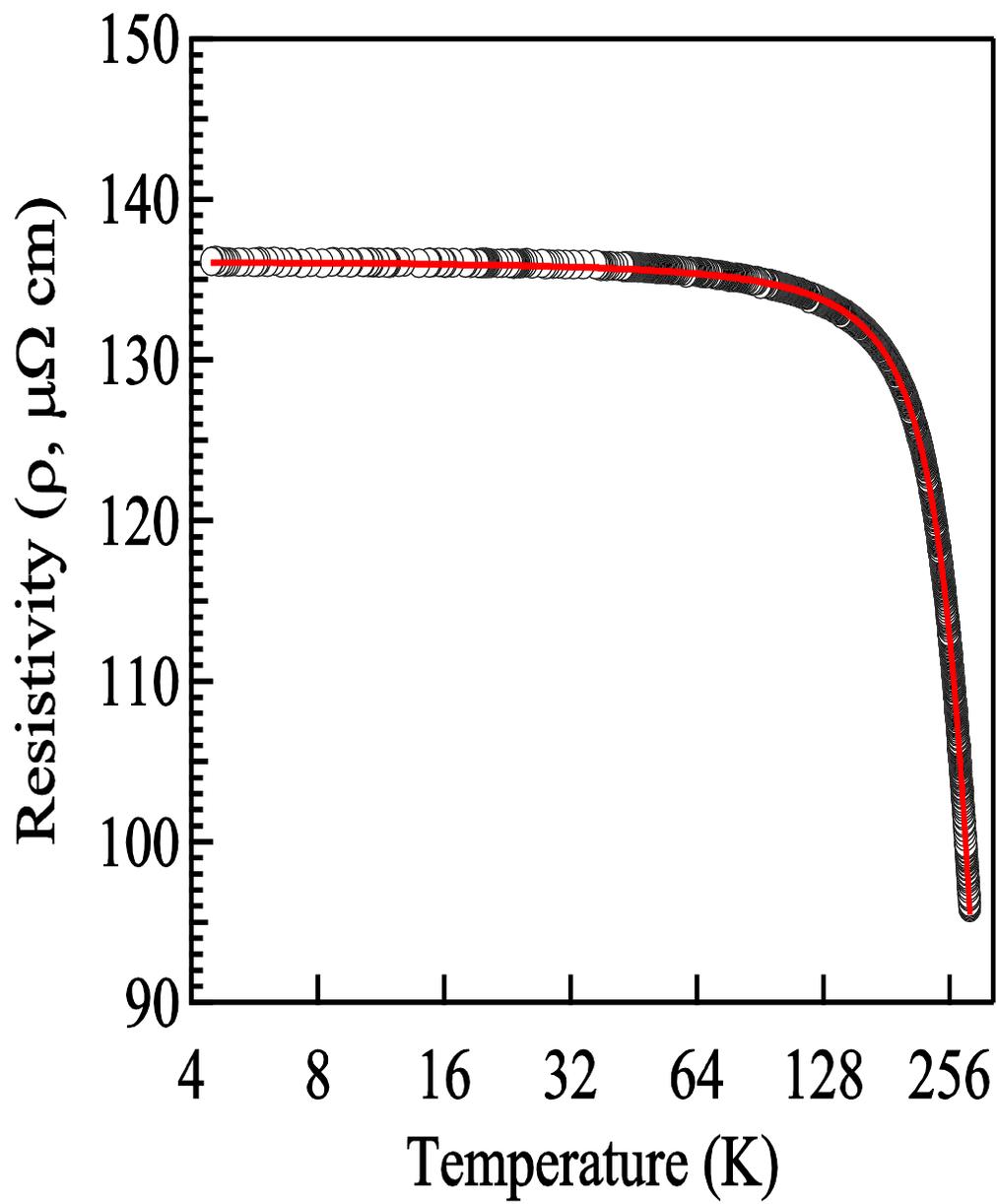

Fig.4.